\documentclass[a4paper,fleqn,usenatbib]{mnras}
\usepackage[T1]{fontenc}
\usepackage{ae,aecompl} 

\usepackage{graphicx}   
\usepackage{amsmath}    
\usepackage{amssymb}    

\title[Very Thin Disc Galaxies]{Very Thin Disc Galaxies in The SDSS Catalog of Edge-on Galaxies}

\author[Bizyaev et al.]{
D. V. Bizyaev$^{1,2}$\thanks{E-mail: dmbiz@apo.nmsu.edu},
S. J. Kautsch$^{3}$,
N. Ya. Sotnikova$^{4}$,
V. P. Reshetnikov$^{4}$,
\newauthor and A. V. Mosenkov$^{5,4,6}$ 
\\
$^{1}$Apache Point Observatory and New Mexico State University, Sunspot, NM, 88349, USA\\
$^{2}$Sternberg Astronomical Institute, Moscow State University, Moscow, Russia\\
$^{3}$Nova Southeastern University, Fort Lauderdale, FL, 33314, USA\\
$^{4}$St.Petersburg State University, 7/9 Universitetskaya nab., St.Petersburg, 199034 Russia\\
$^{5}$Sterrenkundig Observatorium, Universiteit Gent, Krijgslaan 281 S9,
B-9000 Gent, Belgium\\
$^{6}$Central Astronomical Observatory, Russian Academy of Sciences, 65/1 
Pulkovskoye chaussee, St.Petersburg, 196140 Russia
}
 
\begin{document}
 
\date{Accepted XXX. Received YYY; in original form ZZZ}

\pubyear{2016}

\label{firstpage}
\pagerange{\pageref{firstpage}--\pageref{lastpage}}
\maketitle

\begin{abstract}

We study the properties of galaxies with very thin discs using a sample of
85 objects whose stellar disc radial-to-vertical scale ratio determined from
photometric decomposition, exceeds nine.  We present evidences of
similarities between the very thin disc galaxies (VTD galaxies) and low
surface brightness (LSB) disc galaxies, and conclude that both small and
giant LSB galaxies may reveal themselves as VTD, edge-on galaxies.  Our VTD
galaxies are mostly bulgeless, and those with large radial scale length tend
to have redder colors.  We performed spectral observations of 22 VTD
galaxies with the Dual Imaging Spectrograph on the 3.5m telescope at the
Apache Point Observatory.  The spectra with good resolution (R $\sim$ 5000)
allow us to determine the distance and the ionized gas rotation curve
maximum for the galaxies.  Our VTD galaxies have low dust content, in
contrast to regular disc galaxies.  Apparently, VTD galaxies reside in
specific cosmological low-density environments and tend to have less
connection with filaments. Comparing a
toy model that assumes marginally low star formation in galactic discs with
obtained gas kinematics data, we conclude that there is a threshold central
surface density of about 88 $M_{\odot}/pc^2$, which we observe in the case
of very thin, rotationally supported galactic discs.

\end{abstract}


\section{Introduction}

Extremely thin galaxies with large major-to-minor axes ratio 
$a/b$ (typically, determined directly from images) have been noticed in large catalogs of edge-on galaxies 
since a long time ago, e.g., \citep{VV67,FGC,kautsch06}. Edge-on galaxies with large axis ratio, e.g. $a/b > 9 $, 
are often called superthin galaxies \citep{goadroberts79,goadroberts81}. 
Only a few superthin galaxies have been studied thoroughly, such as 
UGC~7321 
\citep{matthews99,matthews00,matthews01,uson03}. A few more large superthin galaxies were studied spectroscopically
\citep{goadroberts81}. Several spiral galaxies with large $a/b$ ratio and without bulges were observed in HI  \citep{matthews00b}.
A large sample of flat and superthin disc galaxies was analyzed and reviewed by \citet{kautsch09} 
using data of optical photometry.
The axis ratio $a/b$ used for the superthin selection in the previous studies was often estimated using faint outer isophotes
in images, sometimes visually, making the $a/b$ axial ratio dependent of used photometric depth
of the images \citep{FGC,rfgc}. The axis ratio estimated this way also can be biased by the presence of a bulge. 
In this paper we use the radial-to-vertical scale ratio ($h/z_0$), which was 
automatically determined for a large sample of verified edge-on galaxies 
by~\citet{EGIS} ("Edge-on disk Galaxies In SDSS", hereafter EGIS\footnote{http://users.apo.nmsu.edu/$\sim$dmbiz/EGIS/}). 
Since some 
large low surface brightness (LSB) galaxies with noticeable bulges are suspected to have thin LSB discs, we do not limit this study to 
bulgeless galaxies only. Instead, we call a galaxy as VTD, i.e. having a very thin disc,  
if its stellar disc's radial-to-vertical scale ratio $h/z_0$ is greater than 9, independent of the bulge contribution. 
Below we outline general properties of our sample of VTD galaxies,
consider their connection to cosmological environment, and estimate properties of their very thin discs, which may help understand
the marginal star formation conditions necessary to form disc galaxies. 
The cosmological framework adopted throughout this paper 
is $H_0$ = 72 km s$^{-1}$ Mpc$^{-1}$ , $\Omega_m$ = 0.3, and $\Omega_\Lambda$ = 0.7.

\section{The Sample of VTD Edge-on Galaxies}

The large sample of verified edge-on galaxies collected by EGIS \citep{EGIS} allows us to select 
a subsample of VTD galaxies large enough for statistical analysis.  
A disadvantage of the EGIS analysis is using a fixed, average gaussian 
Point Spread Function (PSF) for all galaxies. At the same time, using correct PSF is
important in the case of edge-on galaxies, especially for the objects whose
thickness is comparable to the PSF size. We incorporated the corresponding SDSS PSF 
information to the EGIS data release,
and re-ran the photometric profile analysis pipeline with realistic PSFs. 
As a result, we selected 82 galaxies with $h/z_0 >$ 9 (measured in the $r$-band).

The 1D analysis underestimates the radial-to-vertical scale ratio $h/z_0$ in the presence of a 
significant bulge: the scale ratio is biased 10\% or more if the bulge-to-total ratio is greater than 0.4
\citep{mosenkov15}. In order to find possible missing galaxies with large bulges and very thin discs,
we formed a subsample of objects with large $h/z_0$ ratios, which are based on the 3D analysis from EGIS.
Some of those galaxies have been selected by our 1D approach. The images of the rest of the objects were 
visually inspected. The objects with noticeable curvature of the galactic midplane, or with small-scale structural
elements that can confuse the 3D analysis (e.g. edge-on rings or spiral arms)
were removed from the sample. 

The resulting sample comprised 14 galaxies with large bulges and visually thin discs. All the galaxies
were processed through the 1D EGIS pipeline again, while the region for the analysis was adjusted to 
avoid effects of the large bulges. Also, since we expect that stellar and the dust discs in VTD galaxies
have comparable thickness \citep{dalcanton04}, in which the 3D structure analysis 
performed by EGIS was not reliable, the thickness of the dust layer was set equal to that of the stellar
disc, and the 3D pipeline was re-run for the selected group of objects. Neither of these objects showed 
large enough $h/z_0$ to be qualified as VTD, so we left the subsample of 82 VTD galaxies unmodified. 
A well-studied "prototype" superthin galaxy UGC~7321, and two more large thin galaxies 
IC~2233 and UGC~711, were added to the sample for comparison purposes, 
despite their formally determined inverse scale ratio $h/z_0$ is less than nine. Our final
sample comprises 85 VTD galaxies.  

Our sample is presented in Table~1, which shows
EGIS name, radial and vertical scales in kpc, the scale ratio, and the quality flag. The latter indicates 'Y'
if the radial velocity of the galaxy was within the reasonable range (between 0 and 30,000 km/s),
and also if the galactic color $(r-i)$ was reported in EGIS. The flag 'N' indicates that the spatial scales
are incorrect, whereas the scale ratios calculated from the angular values of the scales are right.  
Note that the objects with wrong radial velocities have zeros in the scale columns in Table~1. 

\begin{table*}
\centering
\caption{Selected Sample of Very Thin Disc Galaxies}
\begin{tabular}{lrcccc}
\hline
 EGIS Name & $h$,kpc & $z_0$,kpc & $h/z_0$ & (r-i)$_0$ & Quality \\
\hline
EON\_2.694\_-0.892    & 10.01 &  0.81 & 11.0 &  0.315 & Y \\
EON\_6.305\_13.538    &  0.00 &  0.00 & 10.5 &  0.369 & N \\
EON\_7.069\_24.845    &  0.00 &  0.00 & 11.4 &  0.403 & N \\
EON\_8.366\_-11.103   & 16.66 &  0.81 & 23.6 &  0.336 & Y \\
EON\_8.452\_-9.540    &  0.00 &  0.00 & 11.3 &  0.383 & N \\
\multicolumn{6}{l}{...}\\
\multicolumn{6}{l}{The table is published in its entirety in the electronic edition.}\\
\hline
\end{tabular}
\setcounter{table}{1}\\
The columns show the galaxy name according to the EGIS catalog (which contains coarse decimal RA and Dec 
coordinates), the radial scale length, the vertical scale height, 
the scale ratio, integral color, and the quality flag (see text). \\
\end{table*}

\subsection{Spectral Observations of Very Thin Disc Galaxies}

Systematic study of the properties of VTD galaxies is a difficult task 
without dedicated spectroscopic observations. Just a few galaxies from our
sample have information about their maximum of rotation curve in the HyperLeda 
database. We find that only 63\% of VTD galaxies from our sample 
have radial velocities reported by HyperLeda\footnote{http://leda.univ-lyon1.fr} or SDSS\footnote{http://skyserver.sdss.org}.

We observed a sample of 24 objects with the Dual Imaging Spectrograph
(DIS) on the 3.5m telescope at the Apache Point Observatory (APO).
The observing time was issued in half-night blocks, and between
December 2014 and August 2015 we were granted 8 such observing blocks.
We observed in a high-resolution mode (B1200/R1200 grating),
which provides the spectra resolution of about 5000. In many 
cases we detect major emission lines, other than H$\alpha$, available
in our optical range (H$\beta$, [OIII]4959,5007\AA, [NII]6548,6583\AA, \& [SII]6713,6731\AA).

Typically, each galaxy was observed with 3 exposures from 5 to 20 minutes long.
During the observing nights we obtained a set of biases and dome flats.
A Helium-Neon-Argone wavelength calibration lamp was observed immediately
after each galaxy at the same position in the sky. Spectrophotometric 
standard stars were observed every night.
The data reduction was performed with IRAF standard tools, including
bias subtraction, flat fielding, wavelength calibration, sky line
and sky background subtraction, cosmic ray removal, and flux calibration.
We estimated the heliocentric radial velocity of the galactic 
centre and the maximum rotation velocity $V$ from the spectra for each galaxy. We used the
$H\alpha$ emission line for it,  except for two galaxies in which we did not 
detect any emission lines. In those cases we used the weak NaD absorption lines
at 5893\AA ~and assumed that we estimate the lower limit of the maximum 
rotational velocity from it.

The typical accuracy of the radial velocity and $V$ is 15 and 10 km/s, respectively,
except for the galaxies without the emission lines, where
only a lower limit constraint on the maximum rotational velocity is available. 
Fourteen galaxies from Table~2
have radial velocities reported by the NED database\footnote{https://ned.ipac.caltech.edu}. The mean difference between
the NED and our radial velocities is 9 km/s, and the r.m.s. is 23 km/s for this sample
of fourteen galaxies.  
Table~2 shows the object name (in the EGIS catalog), date of observations, total exposure time, 
our heliocentric radial velocity, and the amplitude of the rotation curve.

\begin{table*}
\centering
\caption{Spectral Observations of Very Thin Disc Galaxies}
\begin{tabular}{llccc}
\hline
Object & Date & Exposure & RV & V$_{max}$ \\
            &         &      min             &    km/s              &  km/s          \\
\hline
EON\_2.694\_-0.892    &      11 Aug 2015  & 35  &  11452 & 152 \\
EON\_17.154\_1.641   &      21 Jun 2015  & 15  &    1964 &  79 \\
EON\_17.275\_19.605 &      16 Aug 2015  & 45  &  12498 &  96 \\
EON\_19.768\_-0.139  &      16 Aug 2015  & 35  &    5230 & 141 \\
EON\_31.087\_6.852   &      11 Aug 2015  & 40  &  23516 & 191 \\
EON\_44.216\_5.686   &      16 Aug 2015  & 45  &   11150 & 132 \\ 
EON\_65.268\_18.389 &    14 Dec 2014  & 20 &   16175 & 232 \\
EON\_119.500\_17.903 &  14 Mar 2015    & 60 & 45001 & 334$^{b}$\\
EON\_126.544\_1.747 &    14 Mar 2015    & 60 & 16773 & 162 \\
EON\_132.574\_3.497 &    14 Mar 2015   &  45 &   8519 & 298 \\
EON\_175.741\_9.394 &    21 Mar 2015   &  50 & 25275 & 272 \\
EON\_194.909\_6.446 &    21 Mar 2015   & 45 &    6363 & 107 \\
EON\_200.709\_19.691$^{a}$ &   21 Mar 2015  &  35 &   6733 & 194 \\
EON\_205.922\_54.952 &  14 Dec 2014   & 40 &  19734 & 317 \\
EON\_211.189\_2.896 &     21 Mar 2015   & 50 & 49931 & 205$^{c}$ \\
EON\_233.587\_57.953 &   21 Mar 2015  & 35 &  29628 & 301 \\
EON\_237.173\_21.870 $^{a}$&   21 May 2015 & 45 &    2160 & 197 \\
EON\_253.776\_39.578 &   21 May 2015 & 45 &  20885 & 213 \\
EON\_266.042\_55.180 &   21 Jun 2015  & 35 &  22872 & 174 \\
EON\_267.275\_64.367 &  18 May 2015  & 55  & 16244 & 193 \\
EON\_310.213\_-6.850 &   21 May 2015  & 50  &  8555 & 113$^{d}$ \\
EON\_332.464\_7.430 &    21 Jun 2015   & 40 &    3955 &  92 \\
EON\_341.857\_-1.255 &   16 Aug 2015  & 50  &  26184 & 202 \\ 
EON\_344.455\_14.189 &  21 Jun 2015   & 45 &  25896 & 215 \\
\hline
\end{tabular}
\setcounter{table}{2}\\
$^a$ Not a VTD galaxy, the $h/z_0$ ratio is close to 6. Used for comparison purposes. \\
$^{b,c}$ No H$\alpha$ line in the spectrum. A line identified as NaD absorption is used. \\
$^{d}$ Raising rotation curve with ring-like features.\\
\end{table*}

\section{General Properties of Our Very Thin Disc Galaxies}

\subsection{Structural Parameters}

\begin{figure}
\includegraphics[width=\columnwidth]{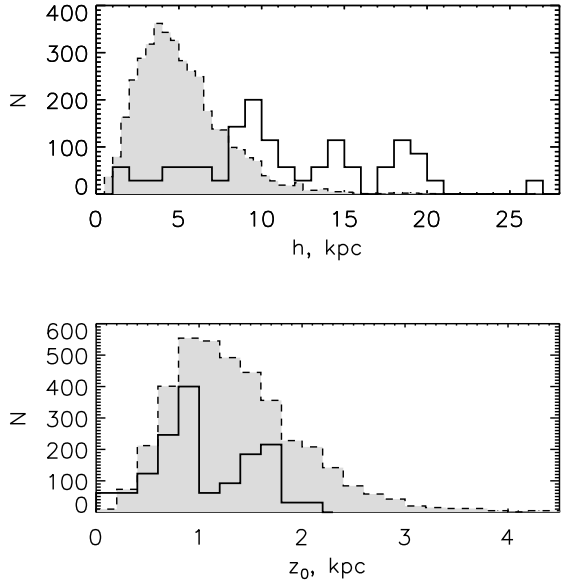}
\caption{The radial scale length (top) and the vertical scale height (bottom) of all EGIS galaxies (dashed curve and grey shaded 
histohram) and of 
VTD subsample (solid curve). The histogram for VTD galaxies is arbitrary scaled in the Y-direction in order to better show the 
plot  (the tallest bins have 6 and 13 objects in the top and bottom panels, respectively). 
\label{fig1}}
\end{figure}

The subsample of VTD galaxies is different in both the stellar disc scale length and 
the stellar disc scale height in comparison to the main sample from EGIS. Figure~\ref{fig1} shows 
that the scale length of VTD discs spans a wide range and is 
biased towards larger values (h = 11.6 $\pm$ 5.6 kpc) when compared to the main sample (h = 5.2 $\pm$ 2.7 kpc). 
The disc scale height of very thin discs is shorter ($z_0$ = 1.0 $\pm$ 0.6 kpc) than
that in the main sample ($z_0$ = 1.3 $\pm$ 0.5 kpc). The mean VTD-to-normal scale ratio is 2.2 for the scale length and 0.71 for the 
scale height. 

\begin{figure}
\centering
\includegraphics[width=6cm]{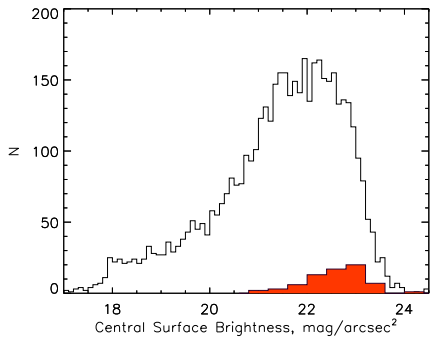}
\caption{The central face-on surface brightness in the $r$-band for all EGIS galaxies (open histogram) and for the 
VTD subsample (filled histogram). The VTD galaxies are mostly LSB galaxies with about 1.5 mag dimmer
surface brightness than regular galaxies in the EGIS catalog.  
\label{fig2}}
\end{figure}

The face-on central surface brightness of VTD discs is biased towards fainter values, see 
Figure~\ref{fig2}. The 
systematic difference of about 1.5 mag between all EGIS galaxies and VTDs. This suggests that the latter are mostly 
LSB galaxies, which is confirmed in detailed studies of UGC~7321 \citep{matthews99,matthews00}. 
Note that the interpretation of Figure~\ref{fig2} is not straightforward because dust effects can systematically move the galaxies 
along the X-axis, and the galaxies with thin stellar discs potentially could be more affected by the dust 
attenuation than galaxies with thicker discs.

\begin{figure}
\includegraphics[width=\columnwidth]{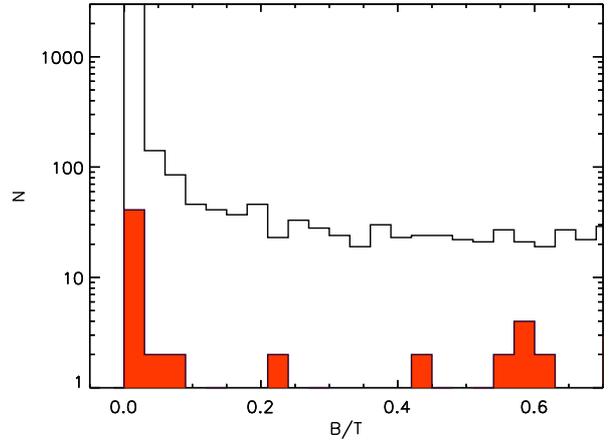}
\caption{The bulge-to-total ratio in the main EGIS sample (open histogram) in the comparison with the VTD subsample
(filled histogram).
\label{fig3}}
\end{figure}

Since we do not constrain our sample by bulgeless galaxies only, we do see some bulges in our VTD objects, although 
they are not large. As follows from Figure~\ref{fig3}, only a small fraction of our VTD sample members has 
noticeable bulges. This conclusion may be biased because the 1D analysis
often reports larger than real scale height in the case of very large
bulge \citep{mosenkov15}. Thus we might systematically
miss galaxies with very large bulges and VTD discs, mostly because of dust effects.
Near-infrared imaging of a statistically significant sample of thin edge-on galaxies
will allow us to update the sample of VTD galaxies in the near future. 

\subsection{Color, Size and Extinction}

\begin{figure}
\includegraphics[width=\columnwidth]{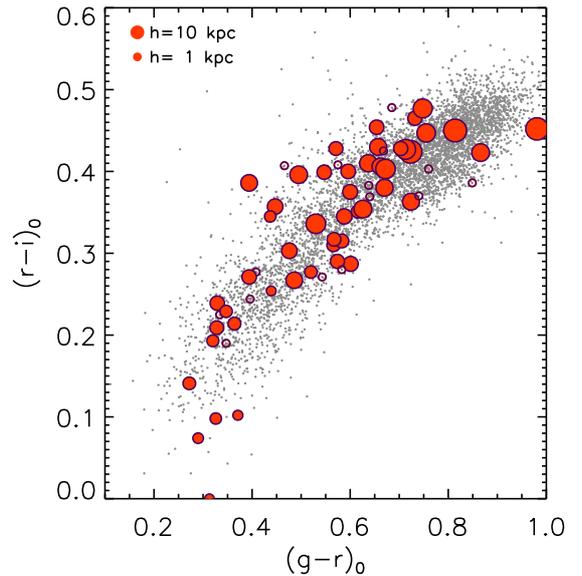}
\caption{The SDSS colors $(g-r)_0$ and $(r-i)_0$ of the main EGIS sample (grey) and of the VTD galaxies (red bullets). 
The colors are corrected for the reddening in the Milky Way, but not corrected for the internal
extinction. The symbol size designates the galaxy scale length. 
The open circles denote the galaxies with unknown redshift and physical size. 
\label{fig5}}
\end{figure}

The broad-band colors of our VTD galaxies do not differ significantly from the main EGIS sample
in the color-color diagram, see Figure~\ref{fig5}. 
The colors of VTD galaxies span the range
typical for the regular galaxies.  
As it is seen in Figure~\ref{fig5}, red VTD galaxies tend to have larger scale length.
The galaxies in the blue corner of the diagram have shorter scalelength, whereas  
the smaller galaxies occupy mostly the lower left corner of the $(g-r)_0$ - $(r-i)_0$ diagram.
The scale length of the VTD galaxies (ranges from 1.6 to 28 kpc)
is denoted by the symbol size in Figure~\ref{fig5}.  
It may suggest that the selected sample of VTD galaxies 
is a mix of different types of objects, including those blue and underevolved, similar to 
UGC~7321,
and large red LSB galaxies like Malin~2 or other large LSB systems \citep{beijersbergen99,oneil00}. 
Note that the colors are corrected for the 
reddening in our Galaxy using \citet{schlegel98} maps, but are not corrected for internal extinction. 

\begin{figure}
\includegraphics[width=\columnwidth]{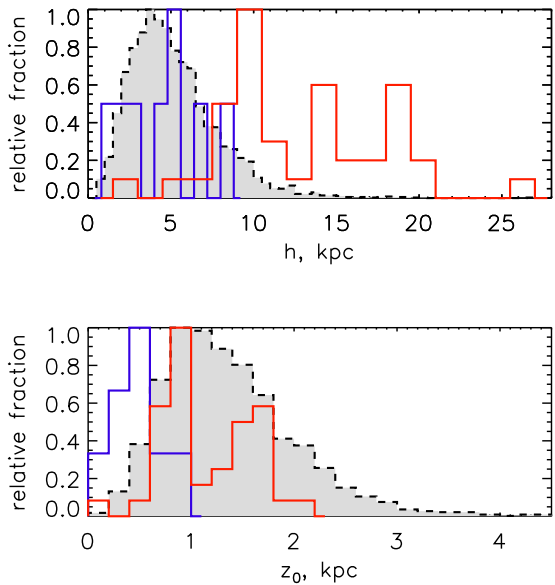}
\caption{The relative fraction of the $h$ and $z_0$ is shown for the same samples as in Figure~\ref{fig1},
 with the VTD galaxies subdivided by the blue and red subsamples (see text). 
The grey filled histogram shows regular EGIS galaxies. The blue and red subsamples are designated by 
the blue and red colors, respectively.
\label{fig60}}
\end{figure}

We split the sample of our VTD galaxies into blue and red subsamples (divided by the color $(r-i)_0$=0.23 mag),
and redraw the first figure in the paper. Figure~\ref{fig60} shows the distribution of the scale length $h$ and scale height $z_0$ of
the VTD galaxies for the separate "blue" and "red" subsamples.   

\begin{figure}
\includegraphics[width=\columnwidth]{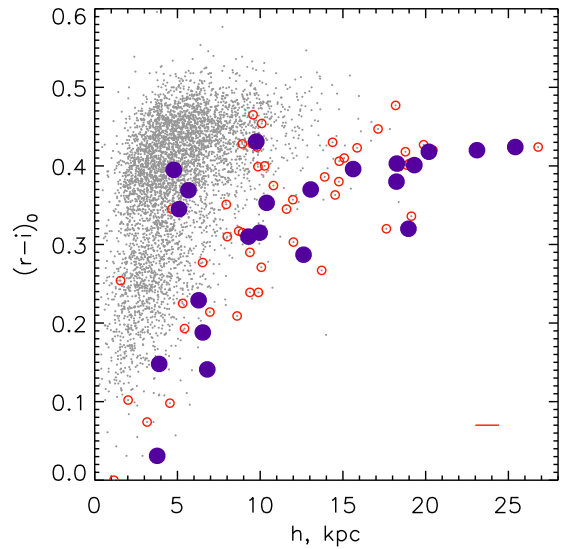}
\caption{The integral (r-i)$_0$ color of EGIS galaxies (grey) in the comparison with their radial scale length expressed in kpc. 
VTD galaxies from our spectroscopic sample with detected H$\alpha$ emission
are designated by the blue bullets. The red circles mark all
other VTD galaxies from our list in Table~2. 
The red bar in the lower right corner shows the average uncertainty of the radial scale length
for the sample of VTD galaxies.
The colors are corrected for the reddening in the Milky Way.
\label{fig6}}
\end{figure}

The VTD galaxies are not typical, as it is seen from their location on the color-size diagram in 
Figure~\ref{fig6}. 
They have bluer color, on average, for the same size as regular EGIS galaxies. The colors of very thin
galaxies can be partly or completely attributed to the low dust extinction in them \citep[][see also discussion below]{maclachlan11},
as well as to underevolved stellar population in LSB galaxies \citep{vorobyov09}. The higher extinction in the 
regular edge-on galaxies can help explain the color difference between the normal and VTD objects  in Figure~\ref{fig6}.
It is worth noting that Figure~\ref{fig6} is qualitatively
similar to the size-color plot from \citet{beijersbergen99} made for LSB galaxies, which also supports the analogy
between the VTD and LSB galaxies.

\begin{figure}
\includegraphics[width=\columnwidth]{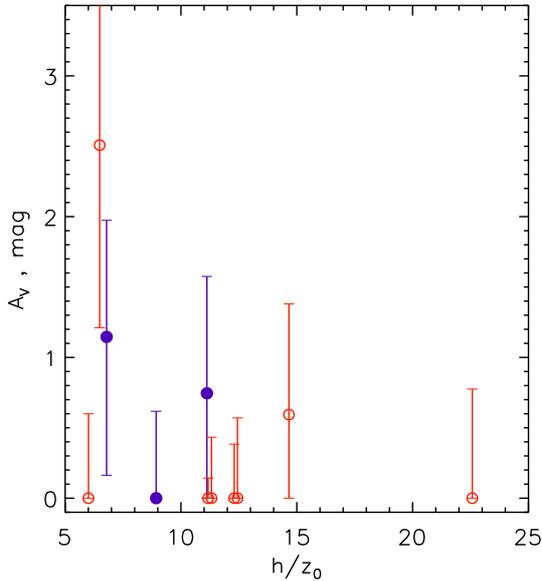}
\caption{The overall dust extinction A$_V$ in the V band estimated from the H$\alpha$/H$\beta$ emission flux ratio
for those 11 galaxies in which the H$\alpha$ and H$\beta$ fluxes were available from our spectroscopy. 
The blue bullets designate the galaxies with $V_{max} ~ \le$ 110 km/s, the red circles show larger galaxies.  
The A$_V$ is corrected for the reddening in the Milky Way.
\label{fig6a}}
\end{figure}

The internal extinction is low in LSB galaxies \citep{matthews01}. \citet{maclachlan11} came to a conclusion that small 
and thin LSB galaxies have low overall dust extinction. Their face-on optical depth is much less than unity. 
The dust scale height in those galaxies is comparable to their stellar scale height. 
By analogy, we expect that our VTD galaxies also have low 
dust extinction. While the H$\alpha$ line is detected in the majority of our objects from Table~2, only 11 galaxies have measurable
H$\beta$ lines. We integrated the hydrogen line fluxes all over the galaxies in order to increase the signal-to-noise and 
estimate the overall extinction from the H$\alpha$/H$\beta$ ratio \citep{charlot01}. We noticed that the background stellar population 
spectrum is not detected in majority of the galaxies or looks extremely weak, 
and the correction for the underlying absorption does not affect the estimated fluxes. 
Figure~\ref{fig6a} shows the overall
extinction A$_V$ corrected for the foreground Milky Way reddening \citep{schlegel98} plotted for the galaxies with 
different thickness. We also include two regular EGIS galaxies (EON\_200.709\_19.691 and EON\_237.173\_21.870) 
that did not meet the "very thin disc" threshold. The bullets designate the 
small galaxies with low amplitude of rotation curve, $V_{max} ~ \le$ 110 km/s. The circles show the galaxies with
$V_{max} ~ >$ 110 km/s.  The VTD galaxies in Figure~\ref{fig6a} indicate low dust extinction, which suggests their
low dust content \citep[see also][]{matthews01,maclachlan11}. 

\section{Very Thin Disc Galaxies in Cosmological Context}

Cosmological simulations suggest that the formation of LSB galaxies may require special cosmological 
conditions: in general, LSB galaxies are born in 15\% less concentrated dark matter halos
than high surface brightness galaxies \citep[e.g.,][]{maccio07}. 
\citet{karachentsev16} found that very thin galaxies tend to have smaller number of satellites than
regular spiral galaxies. 
We investigate if VTD galaxies correlate with elements of the 
large-scale structure, such as voids, superclusters, and filaments.

A catalog of filaments identified in SDSS data by \citet{tempel14} allows us to find the minimum distance
from a galaxy in the EGIS catalog to a filament. We filtered the EGIS catalog and left only the objects
within the same redshift range as our sample of VTD galaxies.We also limited the maximum radial velocity
in both samples by 25,000 km/s since the majority of VTD galaxies are have lower radial velocities. 
Note that the conclusions presented below do not change if the maximum radial velocity is limited by 
slightly different value (e.g. 30,000 km/s.)
In this case the Kolmogorov-Smirnov test suggests that the
probability that the distances in the two subsamples are drawn from the same distribution equals to 0.77. 
Then we estimated three-dimensional  
distances to the filaments from \citet{tempel14} for the objects from both samples.

\begin{figure}
\includegraphics[width=\columnwidth]{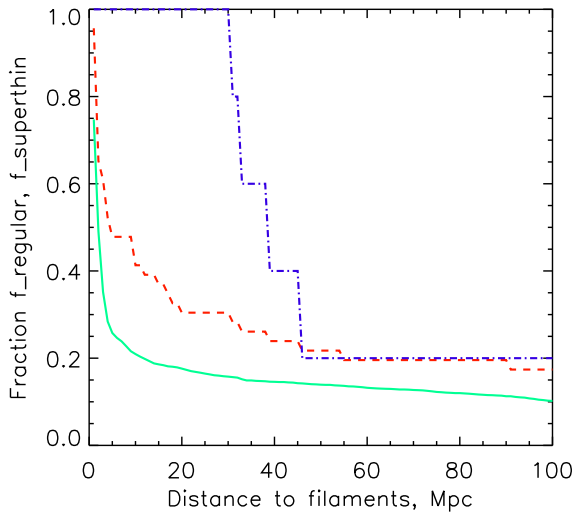}
\caption{The fraction of edge-on galaxies that reside farther than certain distance from filaments. 
The solid green curve designates regular EGIS galaxies, while the dashed red curve shows our VTD galaxies. 
The dash-dotted blue curve denotes our small VTD galaxies with h $<$ 4.5 kpc. 
The solid and dashed lines are different, which indicates that the fraction of VTD galaxies that 
associate with filaments is twice as less than that of the regular (non-thin disc) edge-on galaxies. 
\label{fig7}}
\end{figure}

We observe significant difference in how  
the VTD and regular galaxies associate with the filaments. 
Figure~\ref{fig7} shows the fraction of galaxies that are farther than certain distance from filaments
for three groups of objects: all non-thin disc galaxies, all VTD galaxies, and those from the latter group whose
scale length is shorter than 4.5 kpc. Here the fraction that equals to 1 means that there are no objects
closer than this distance from the filament, and the fraction of 0 means that all objects of this kind
are located closer than certain distance to the nearby filament. 

It is seen that within a certain distance from filaments we observe a smaller fraction of VTD galaxies than 
regular EGIS objects. The fraction of non-associated supethins is more than twice as large than 
that of the regular edge-on galaxies starting with a few Mpc from filaments. 

Interesting to notice that 
LSB galaxies also tend to avoid the inner regions in filaments ($\sim$ inner 5 Mpc), and prefer to reside in the 
outer regions there \citep{mo94,rosenbaum04}.  Since large and small galaxies tend to have different
colors (Figure~\ref{fig5}), which may reveal different evolution scenarios, we also check if the radial scale length 
affects the fraction of the galaxies associated with filaments. The dash-dotted blue curve in 
Figure~\ref{fig7}
designates relatively small VTD galaxies defined as those with the scalelength h $\le$ 4.5 kpc.  
This subsample of small VTD galaxies indicates higher than in the general VTD subsample fraction 
of objects that do not associate with filaments.
In general, we observe that VTD galaxies tend to reside in low-galaxy density environments, 
similar to completely bulgeless disc galaxies \citep{kautsch2009}.

The smaller VTDs are apparently, underevolved objects that were born and evolved 
in unperturbed environment, with slower rate of accretion from filaments.  
This, probably, explains why we can find such galaxies at larger distances 
from filaments with respect to all other objects in our sample.

In order to estimate the dependence of our conclusions of the sample selection, 
we varied the lower and upper thresholds of the radial velocity that determines the inclusion
of objects into both samples. We found that the conclusion of this section
that VTD galaxies tend to avoid filaments (in the comparison with the regular galaxies) holds for a wide variety
of the distance-limited subsamples. The conclusion is independent of the estimated 
Kolmogorov-Smirnov (KS) probability (that is the probability that the samples of the regular and VTD galaxies 
are drawn from the same parental population).
The KS probability can be as high as 0.98 for certain cases of the radial velocity thresholds selection.

\begin{figure}
\includegraphics[width=\columnwidth]{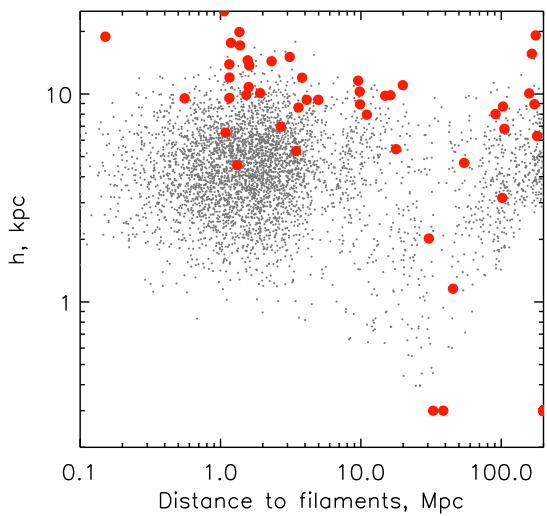}
\caption{The distance to filaments versus the scale length of the galaxy. The sample of regular, non VTD
galaxies is designated by grey dots. The VTD galaxies are marked with the red bullets. 
\label{fig777}}
\end{figure}

Figure~\ref{fig777} shows both VTD and regular EGIS samples of edge-on galaxies on the distance-to-filament versus
the scale length diagram. Figure~\ref{fig777} highlights that small VTD galaxies can be found at larger distances from 
filaments, whereas the proximity to the filaments of the large VTDs resembles that of the regular EGIS galaxies.  

We observe no difference between the correlation of all EGIS galaxies and VTD galaxies with voids: both 
groups of galaxies stay away from voids, according to the catalog by \citet{varela12}. 
We checked if there is a difference in the fraction of regular and VTD galaxies
that associate with superclusters identified by \citet{liivamagi12}, and did not find significant difference.
Figure similar to Figure~\ref{fig7} made with the distance to superclusters instead of filaments shows no
difference between all EGIS and VTD galaxies.

\section{How to Make a Very Thin Disc Galaxy?}

The main ingredients of known VTD galaxies are massive dark matter halos and LSB stellar discs
\citep{matthews99,uson03,bm02,bizyaev04,sotnikova06,bm09,karachentsev16}.
The gas fraction is also high in some VTD  galaxies \citep{matthews00b,uson03,banerjee09}.
It was found that the gravitational potential of dark matter halo dominates at all radii in
a VTD galaxy UGC~7321 \citep{banerjee09}.

\subsection{Properties of Dark Matter Halos in Very Thin Disc Galaxies}

A massive dark halo is a necessary but not sufficient 
condition: \citet{saha14} found that a galaxy has to avoid forming 
a bar or prominent spirals during its evolution to be a VTD. 
\citet{banerjee13} analyzed the halo concentration parameter in two galaxies 
and concluded that a compact dark matter halo with its scale less
the radial scalelength of the stellar disc is required to assemble 
a VTD galaxy. 
\citet{sotnikova05} concluded that the presence 
of a compact, not necessarily massive bulge in a spiral galaxy may 
be enough to suppress the bending instability and to keep the stellar 
disc very thin. This finding supports the conclusions by
\citet{banerjee13}, because \citet{sotnikova05} noticed that 
the gravitational potential of a compact bulge has the same 
effect as the potential of a compact dark halo.
\citet{kh10} simultaneously modeled rotation curves and thickness 
of several galaxies, including three very thin ones. 
They found that dark halos dominate by mass in the galaxies 
with very thin stellar discs.
The presence of a compact dark halo 
in galaxies with very thin discs is not confirmed by \citet{kh10}.
Instead, the results of \citet{kh10} indicate that stellar discs must 
have a low surface density to keep them very thin. 

An interesting example comes from the dynamical analysis of components of 
a giant LSB galaxy Malin~2: \citet{kasparova14} confirm a low 
surface brightness stellar disc in Malin~2, and also a very massive 
dark halo. The stellar disc thickness estimated from dynamical 
equilibrium reasons is of the order of 1 kpc in Malin~2 
(A. Kasparova, private communication), so the galaxy 
can be considered as a VTD, with a large scale ratio 
$h/z_0 \sim$  20. The halo scale length was estimated by 
\citet{kasparova14}, and the conclusion is opposite to that by 
\citet{banerjee13}: an extended and spread-out halo is required 
to make a giant LSB (and VTD) galaxy. 
Thus we suspect that the halo concentration parameter may work
different ways in the evolution of small and large 
VTD galaxies, but in all cases a low surface density 
stellar disc is required. 

\subsection{Disc Surface Density Threshold and Very Thin Stellar Discs}

The marginal conditions for star formation in galactic discs can be explored and 
quantified with the help of VTD galaxies. VTDs exhibit low-density discs 
and gravitationally dominating dark halos. This brings them close to the minimum conditions, 
which are necessary for starting the star formation. Therefore, VTD galaxies 
provide an additional tool, the disc thickness, which cannot be 
infinitely small. The one-dimensional gas velocity dispersion is 
of the order of 6 km/s in the star forming media 
\citep[see e.g.,][]{kennicutt89}, and the stellar disc formed 
from the gas cannot decrease its vertical velocity dispersion being collisionless.

A toy model of an exponential stellar disc embedded into a spherical 
gravitational potential brings up simple equations to estimate the 
relative disc thickness 
\citep{zasov91, zasov02,kregel05b, sotnikova06,bm09, kh10}. 
We assume that the disc mass is $M_d = 2 \pi \, \Sigma_0 \, h^2$, 
and the total mass of the galaxy within four disc
scale lengths is $M_t = 4\, V^2\, h / G$. 
Here $\Sigma_0$ is the central surface density of the stellar disc,
$V$ is the circular velocity, which can be assumed as constant in 
the case of a flat rotation curve, 
and $G$ is the gravitational constant. 
We consider the discs are in equilibrium in the vertical 
direction, with their vertical scaleheight 
$z_0$ determined via the vertical equilibrium condition for 
the isothermal slab~\citep{spitzer42}: 
$\sigma_z^2 = \pi \, G \, \Sigma(R) \, z_0$, where 
$\sigma_z$ is the vertical stellar or gas velocity dispersion.

The radial stellar velocity dispersion $\sigma_R$ in a disc that is
marginally stable against axisymmetric perturbations in 
its plane, is $\sigma_R = 3.36\, G  \Sigma(R) / \kappa$ \citep{toomre64}, 
where {\it $\kappa$} is the epicyclic frequency. If we take into account 
the non-axisymmetric perturbations and a finite layer thickness, the 
minimum radial dispersion should be greater than 
$\sigma_R \geq Q 3.36\, G  \Sigma(R) / \kappa$, where $Q>1$ is the 
Toomre parameter \citep[e.g.,][and references therein]{polyachenko97}. 
The radial profile of $Q$ usually has 
a wide minimum with the value $Q \approx 1.4$ in the region of 
$(1.5-2) \, h$ in the marginally stable discs \citep[see, e.g., numerical simulations by][]{khoperskov03}.
The value of $ Q \approx 1.4$ is 
consistent with the empirical criterion of the gravitational stability 
\citep{kennicutt89}. 
Thus, we can safely assume that $Q$ is a
constant at intermediate distances to the galactic centre.
The epicyclic frequency at the region of the flat
rotation curve is $\kappa = \sqrt{2} V / R$, and 
we get $z_0 \sim (M_d/M_t) h$. 
At $R = 2\,h$ the total-to-disc mass ratio is 
$M_t / M_d \, \gtrsim \, 1.2 \, (\sigma_z / \sigma_R)^2 \, 
(h / z_0) \, (Q / 1.4)^2$.

The low vertical dispersion in disc galaxies allows us to introduce a few more 
simplifications. Note that if we consider early epochs of the galaxy 
formation, the discs should be assumed mostly gaseous, for which 
$\sigma_R = \pi\, G\, Q\,  \Sigma(R) / \kappa$ \citep{safronov60} 
and $(\sigma_z / \sigma_R)$ = 1. 
In general, considering the instability of multi-component galactic 
discs \citep{jog84,rafikov01,romeo13} as well as finite gas and 
stellar layer thickness \citep[e.g.,][]{bogelman09} should introduce 
corrections to the stability criterion. 
Fortunately, the proximity of the 
vertical velocity dispersion in the stellar and gas discs, as well as the 
small disc thickness, make the pure gas disc stability criterion 
equation well applicable in the case of the VTD discs.
In this case
\begin{equation}
\label{eq1}
M_t / M_d \, \gtrsim \, 1.1\, (h/z_0)\, (Q/1.4)^2 \, .
\end{equation}
Formally, equation~(\ref{eq1}) does not put any constraints on the 
stellar disc thickness and if $M_t / M_d \rightarrow \infty$, 
the ratio $h/z_0$ grows infinitely, too.

Although we considered a purely gaseous disc in our toy model, 
the stellar component inevitably emerges in real galaxies. It is worth 
noticing that if we apply a two-component stability criterion \citep{jog84,efstathiou00} 
in which the stellar velocity dispersion is equal or higher than that in the gas, 
we obtain a higher critical density necessary to make the disc 
(stellar and gaseous together, in this case) unstable. 
The stellar or gas discs can be stable if considered taken apart,
but their combination can be unstable, at the same time. 
In this case our approximation of purely gaseous disc at the beginning, 
when the star formation just started, constrains the lower limit of the critical 
density necessary for starting the disc fragmentation.

One more limitation to the disc thickness comes from the inability to 
make vertical component of the velocity dispersion arbitrary 
small. We assume that one-dimensional gas velocity dispersion is limited by 
(10/$\sqrt{3}$) km/s, and that the stellar population cannot
inherit the vertical velocity dispersion less than this value. 
Starting with equation~(\ref{eq1}), we substitute 
$\sigma_z  = 10 / \sqrt{3}$ km/s,
$\mu = M_t / M_d$, $Q = 1.4$, and $R = 2\,h$ in it, and get
$h/z_0 = 2\, V^2 / (\mu \, e^2)$, where $e$ is the base of the 
natural logarithm. In more convenient designation this can be written
\begin{equation}
\label{eq2}
h/z_0 \lesssim (81/\mu) \, V_{100}^2 \, ,
\end{equation}
where $V_{100} = V / 100$ km/s.
Equation~(\ref{eq2}) predicts very thin discs 
($h/z_0 \lesssim 20$ for $V_{100} \sim 1$) for galaxies with massive dark halos ($\mu \sim 4$) 
\citep{bm09,kh10,uson03,kasparova14}, while the predicted
thickness for $V_{100}\, > 2$ galaxies becomes unrealistic. 
Large values of $\mu = M_t / M_d$ would help partly ``fix'' 
equation~(\ref{eq2}): 
\citet{mosenkov10} report slightly higher maximum values of $\mu$ 
up to 8, but it will not prevent equation~(\ref{eq2}) from displaying 
very high ratios $h/z_0$ for large galaxies with $V_{100} > 2$.

The star formation threshold criteria in the form of a plain 
critical surface density were considered in the past 
\citep[e.g.,][]{guiderdoni87,skillman87,vdhulst87,impey89,phillips90,clark14}.
We consider this for our galaxies as well, because this threshold density can limit the disc surface
density.
We employ a relationship between the radial scale length and 
the rotation velocity $h \sim V^{1.5}$, as follows from 
\citet{courteau07,hall12}.
The coefficient in the equation can be calibrated using existing 
kinematic measurements available for several VTD galaxies 
from our sample, see the next section. 
In this case equation~(\ref{eq1}) can be written
\begin{equation}
\label{eq3}
h/z_0 = 107\, V^{0.5} / \Sigma_{0,c} \, .
\end{equation}
Here $V$ is expressed in km/s, and $\Sigma_{0,c}$ in 
M$_{\odot}$/pc$^2$; $Q = 1.4$, $\sigma_z/\sigma_R = 1$.
In this case we expect 
$h/z_0 \sim 10$ for small galaxies like UGC~7321, and 
$h/z_0 \gtrsim 15$ for giant LSB galaxies like Malin 2 
(given a proper value of $\Sigma_{0,c}$). 

Since equations (\ref{eq2}) and (\ref{eq3}) have different functional 
dependence of the rotation curve amplitude $V$, we can 
distinguish between the two cases considered above. The thinnest 
discs in the VTD galaxies should highlight the cases
of the lowest vertical stellar velocity dispersion and lowest 
stellar density. 


\begin{figure}
\includegraphics[width=\columnwidth]{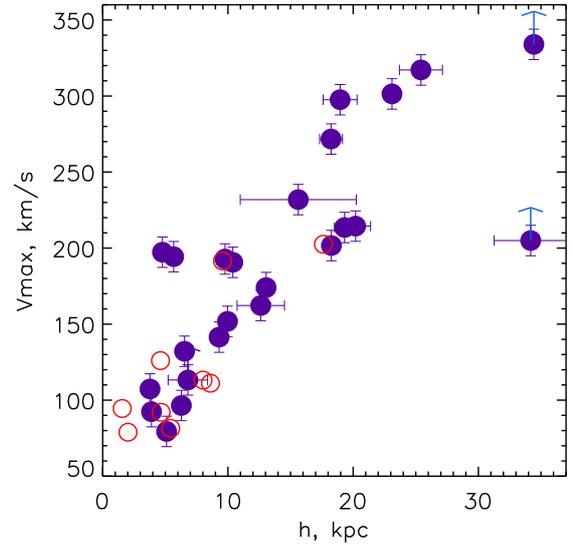}
\caption{The rotation curve maximum $V$ versus the scale length diagram 
for nine galaxies with published data (the red open circles) and for the sample from Table 2 
(the blue bullets). The diagram suggests that the majority of the galaxies
follow the same trend. The three galaxies with the lower limit estimations for $V$ (see text) 
are marked with arrows. 
One of these galaxies should have much higher $V$ than the lower limit. 
\label{fig9}}
\end{figure}

Figure~\ref{fig9} demonstrates the relation between the rotation curve maximum
$V$ and the radial scale length 
$h$ for nine galaxies with data from HYPERLEDA (open circles) and for 
the galaxies observed by us 
(filled circles).  The trend corresponds to the $h \sim V^{1.5}$ dependency. 
The arrows indicate the two cases of the lack of emission lines in spectra, when
NaD absorption line was used to constrain the $V$, and one more case of purely
raising rotation curve with ring-like brightness distribution in the H$\alpha$
rotation curve. The latter indicates that we observe emission from a gas ring in the galaxy,
whereas the outer regions can rotate faster than gas in the ring. 

\subsubsection{The Disc Surface Density Threshold}

\begin{figure}
\includegraphics[width=\columnwidth]{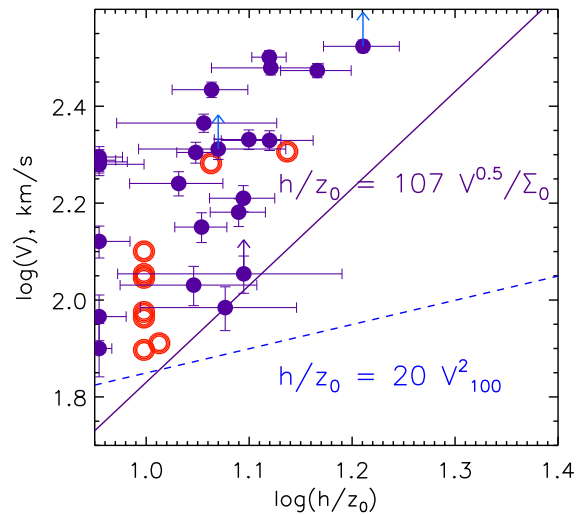}
\caption{The rotation curve maximum $V$ versus the inverse stellar disc thickness 
$h/z_0$. The two lines, solid and dashed, correspond 
to the cases of a surface density threshold, and no threshold, 
respectively. The red open circles show the data available for 
several VTD galaxies from literature. The blue bullets
designate our sample with APO/DIS spectral observations.
\label{fig10}}
\end{figure}

Figure~\ref{fig10} shows the comparison between $h/z_0$ and $V$ for 
our VTD galaxies.
The open circles designate the galaxies with  
the rotation curve maximum $V$ found in literature (HYPERLEDA), 
while the filled circles show our APO/DIS measurements (\S2.1).
The arrows indicate the galaxies with the lower limit of $V$, same as in
Figure~\ref{fig9}.

The solid and dashed lines correspond to equations (\ref{eq3}) 
and (\ref{eq2}) with $\mu = 4$, respectively. 
The observing point allocation in Figure~\ref{fig10} strongly 
favors the case of a star formation threshold surface density. 
Since all galaxies (not only VTDs) are located 
above the solid line because the threshold should prevent them 
from moving below 
this line), the lowest points help determine the ``envelope'' line 
that corresponds to certain lowest  $\Sigma_{0,c}$. 
We determine that the solid curve in Figure~\ref{fig10} corresponds 
to $\Sigma_{0,c}$ = 88 M$_{\odot}$/pc$^2$. As it follows from the assumed 
initial conditions and assumptions for the equations, this is the 
stellar and gas surface density summed together. 
It is interesting to notice that the upper left corner in Fig.~11 in
\citet{matthews00b} demonstrates a similar ``envelope'' feature and 
confirms our finding, although the definition of 
$a/b$ ratio in \citet{matthews00b}  
comes from a visual estimate, therefore it is hard to compare it 
quantitatively to our results. \citet{kregel05} show this ``envelope'' feature in their $h/z_0$ versus $V$ plots as well.
Note that the predictions of our toy model are in qualitative agreement with 
N-body simulations of UGC~7321, which suggest that the central surface density
in the stellar disc ranges between 50 \citep{banerjee13} 
and 200 M$_{\odot}$/pc$^2$ \citep{kh10}.

The galaxies that are located above both the solid and dashed lines in 
Figure~\ref{fig10} have their disc surface density above the 
threshold and the stellar velocity dispersion above the minimum value. 
It does not mean that galaxies with lower surface density and, hence, 
lower surface brightness cannot exist: galaxies can have a lower 
central surface density, but their discs will be thick and have a 
relatively high vertical stellar velocity dispersion, i.e.  
they will be more pressure supported and less rotationally supported. In the latter sense they 
will resemble Irr/Im galaxies rather than Sd. The minimum central 
surface density that we observe in the large, rotationally supported 
disc galaxies, may be a manifestation of the disc formation 
regulating processes via the spin angular momentum parameter 
$\lambda$ \citep{peebles69,dalcanton97}: LSB galaxies
with massive dark halos tend to have large $\lambda$ and a low dark 
matter halo concentration index \citep{maccio07}. Since we assumed 
the marginal stability of galactic discs to evaluate 
equation (\ref{eq1}), this threshold density is an additional 
threshold, independent of the Toomre-Kennicutt large scale 
instability criterion \citep{kennicutt89}. 

Our toy model uses several simplifications and assumptions. 
``Typical'' values of numerical coefficients may not work well 
in all cases of the extreme conditions, which we consider.
The modelling of real galaxy rotation curves 
with the application of additional constraints from the stellar disc thickness
for a larger sample of VTD galaxies should help to better understand 
their dynamical status and the star formation threshold.
The essential transparency of the dust layer in the galaxies is an additional 
feature, which should simplify the estimation of the parameters of the galactic components 
from the rotation curve modelling (e.g., as shown in \citet{zasov03,kregel05b}).

\subsection{Very Thin Disc Galaxies and LSB galaxies}

The properties of VTD galaxies resemble those of LSB disc galaxies. 
Both classes of galaxies have low surface brightness, both are relatively deficient 
in dust, and prefer to reside outside the core filament regions. 
VTD galaxies possess large fraction of dark matter, similar to 
many LSB galaxies. We can conclude that VTD galaxies are mostly 
LSB galaxies, but the opposite is incorrect, in general: not all 
LSB galaxies are VTD, as well as not all LSB galaxies have 
massive dark halos \citep{graham02}. Moreover, extreme LSB
galaxies with a surface density below $\sim$ 90 M$_{\odot}$/pc$^2$
should have moderate (less than nine) radial-to-vertical disc scale ratios
and the stellar velocity dispersion close to that in the galactic gas medium. 
Additional studies of statistically large samples of VTD galaxies are needed to 
verify our conclusions. 

\section{Conclusions}

We select 85 VTD galaxies with large radial-to-vertical scale 
ratios $h/z_0$ from the EGIS catalog of edge-on galaxies \citep{EGIS}. 
The VTD galaxies have larger scale lengths and 
shorter scale heights than regular EGIS objects in general. The objects with large 
radial and vertical scales tend to have redder colors, whereas smaller VTD 
galaxies have bluer colors. 

VTD galaxies are mostly LSB stellar systems with low dust extinction. 
They may possess bulges, but a large fraction of VTD galaxies have very small 
bulges (bulge-to-total luminosity ratio is less than 0.1) or no bulges at all.

VTD galaxies from our sample avoid large-scale filaments twice as frequent than 
regular EGIS objects, thus suggesting that the VTDs are located in more isolated 
environment. At the same time, VTD galaxies possess very massive  
and spread out dark matter halos, which makes the dark-to-luminous mass ratio
several times greater than that in regular spiral 
galaxies. Further studies of  
correlations between the properties of the dark halos around VTD 
galaxies and cosmological structures in which they reside should 
help identify specific cosmological conditions necessary to create 
VTD, dark matter dominated disc galaxies. 

Correlation between the scale ratio $h/z_0$ and maximum rotational velocity,
as well as the lack of very thin and low-massive galaxies, 
suggests that the formation of the disc component in galaxies is 
regulated by a threshold surface density. Using kinematic data 
available for a sample of VTD galaxies observed with 
the Dual Imaging Spectrograph at the Apache Point Observatory,
we conclude that 
the minimum central surface density in the VTD galaxies is 
88 M$_{\odot}$/pc$^2$. Galaxies with less central surface density 
indicate a small $h/z_0$ ratio (i.e. they are not VTD), 
and they are less rotationally supported systems. 

\section*{Acknowledgements}

%

DB is supported by RSF grants RSCF-14-50-00043 (spectra acquisition and
reduction) and RSCF-14-22-00041 (data interpretation and modeling.) AM is a
beneficiary of a mobility grant from the Belgian Federal Science Policy
Office.  We acknowledge partial financial support from the RFBR grants
14-02-00810 and 14-22-03006-ofi.  Based on observations obtained with the Apache Point
Observatory 3.5-meter telescope, which is owned and operated by the
Astrophysical Research Consortium.

We appreciate valuable suggestions by Heinz Andernach (Univ.  Guanajuato)
for his help in increasing our sample of EGIS galaxies with known parameters. 
We thank Dmitry Makarov (SAO RAS) and Vasily Belokurov (University of
Cambridge) for comments on many individual galaxies in the EGIS catalog. We
thank anonymous referee for helpful constructive comments that improved the paper.

We acknowledge the usage of the HyperLeda database.  This research has made
use of the NASA/IPAC Extragalactic Database (NED) which is operated by the
Jet Propulsion Laboratory, California Institute of Technology, under
contract with the National Aeronautics and Space Administration.

{}

\bsp    
\label{lastpage}


\begin{thebibliography}{}

\bibitem[Banerjee et al.(2009)]{banerjee09}Banerjee, A., Jog, C., \& Matthews, L. 2009, ASP Conf. Ser., 407, 99

\bibitem[Banerjee \& Jog(2013)]{banerjee13}Banerjee, A. \& Jog, C. 2013, \mnras, 431, 582

\bibitem[Beijersbergen et al.(1999)]{beijersbergen99}Beijersbergen, M., de Blok, W. J. G., \& van der Hulst, J. M. 1999, \aap, 351, 903

\bibitem[Bizyaev \& Kajsin(2004)]{bizyaev04}Bizyaev, D. \& Kajsin, S. 2004, \apj, 613, 886

\bibitem[Bizyaev \& Mitronova(2002)]{bm02}Bizyaev , D. \& Mitronova, S.  2002, \aap, 100, 200

\bibitem[Bizyaev \& Mitronova(2009)]{bm09}Bizyaev , D. \& Mitronova, S.  2009, \apj, 702, 1567

\bibitem[Bizyaev et al.(2014)]{EGIS}Bizyaev, D., Kautsch, S., Mosenkov, et al. 2014, \apj, 787, 24 (EGIS)

\bibitem[Bogelman \& Shlosman(2009)]{bogelman09} Bogelman, M. \& Shlosman, I. 2009, \apjl, 702L, 5

\bibitem[Bottema(1993)]{bottema93}Bottema, R., 1993, \aap, 275, 16   

\bibitem[Charlot \& Longhetti(2001)]{charlot01} Charlot, S. \& Longhetti, M. \mnras, 323, 887

\bibitem[Clark \& Glover(2014)]{clark14} Clark, P. C. \& Glover, S. 2014, \mnras, 44, 2396

\bibitem[Courteau et al.(2007)]{courteau07}  Courteau, S., Dutton, A., van den Bosch, F., et al. 2007, \apj, 671, 203

\bibitem[Dalcanton et al.(1997)]{dalcanton97}Dalcanton, J. J., Spergel, D. F., \& Summers, F. J. 1997, \apj, 482, 659

\bibitem[Dalcanton et al.(2004)]{dalcanton04}Dalcanton, J. J., Yoachim, P., \& Bernstein, R. A. 2004, \apj, 608, 189

\bibitem[Dehnen \& Binney(1998)]{dehnen98} Dehnen, W. \& Binney, J.J. 1998, \mnras, 298, 387 

\bibitem[Efstathiou(2000)]{efstathiou00} Efstathiou, G. 2000, \apj, 317, 697

\bibitem[Gerssen et al.(2000)]{gerssen00}Gerssen, J., Kuijken, K., \& Merrifield, M. 2000, \mnras, 317, 545

\bibitem[Graham(2002)]{graham02}Graham, A. 2002, \mnras, 334, 721

\bibitem[Goad \& Roberts(1979)]{goadroberts79} Goad, J. W., Roberts, M. S. 1979, \baas, 11, 668

\bibitem[Goad \& Roberts(1981)]{goadroberts81} Goad, J. W., Roberts, M. S. 1981, \apj, 250, 79

\bibitem[Guiderdoni(1987)]{guiderdoni87} Guiderdoni, B. 1987, \aap, 172, 27

\bibitem[Hall et al.(2012)]{hall12} Hall, M., Courteau, S., Dutton, A., et al. 2012, \mnras, 425, 2741



\bibitem[Impey \& Bothun(1989)]{impey89} Impey, C. \& Bothun, G. 1989, \apj, 341, 89

\bibitem[Jarrett et al.(2003)]{2MASS} Jarrett, T. H., Chester, T., Cutri, R., et al. 2003, \aj, 125, 525

\bibitem[Jog \& Solomon(1984)]{jog84} Jog, C. \& Solomon, P. 1984, \apj, 276, 127 

\bibitem[Karachentsev et al.(1993)]{FGC}
Karachentsev, I. D., Karachentseva, V. E., Parnovskij, S. L. 1993, Astronomische Nachtichten, 314, 97

\bibitem[Karachentsev et al.(1999)]{rfgc}
Karachentsev, I. D., Karachentseva, V. E., Kudrya, Yu. N.,
et al., 1999, Bull. Spec. Astr. Obs., 47, 5

\bibitem[Karachentsev et al.(2016)]{karachentsev16}
Karachentsev, I. D., Karachentseva, V. E. \& Kudrya, Yu. N., 2016, arXiv:1605.03734


\bibitem[Kasparova et al.(2014)]{kasparova14}Kasparova, A., Saburova, A., Katkov, I., et al. 2014, \mnras, 437, 3072

\bibitem[Kautsch et al.(2006)]{kautsch06}Kautsch, S. J., Grebel, E. K., Barazza, F. D., \& Gallagher, J. S., III 2006, \aap,  445, 765

\bibitem[Kautsch et al.(2009)]{kautsch2009}Kautsch, S. J., Gallagher, J. S., \& Grebel, E. K. 2009, Astronomische Nachrichten, 330, 1056

\bibitem[Kautsch(2009)]{kautsch09}Kautsch, S. J. 2009, \pasp, 121, 1297

\bibitem[Kennicutt(1989)]{kennicutt89}Kennicutt, R. C. 1989, \apj, 344, 685

\bibitem[Khoperskov et al.(2003)]{khoperskov03} Khoperskov, A., Zasov, A., \& Tyurina, N.  2003, Astron. Reports, 47, 357

\bibitem[Khoperskov et al.(2010)]{kh10} Khoperskov, A., Bizyaev, D., Tiurina, N., \& Butenko, M. 2010,  Astronomische Nachtichten, 331, 731


\bibitem[Kregel et al.(2005)]{kregel05}Kregel, M., van der Kruit, P. C., \& Freeman, K. C. 2005, \mnras, 358, 503

\bibitem[Kregel \& van der Kruit(2005)]{kregel05b}Kregel, M. \& van der Kruit, P. C. 2005, \mnras, 358, 481

\bibitem[Liivamagi et al.(2012)]{liivamagi12}LiivamŠgi, L., Tempel, E., \& Saar, E. 2012, \aap, 539, 80

\bibitem[Maccio et al.(2007)]{maccio07}Maccio, A., Dutton, A., van den Bosch, F., et al. 2007, \mnras, 378, 55

\bibitem[MacLachlan et al.(2011)]{maclachlan11} MacLachlan, J. M., Matthews, L. D., Wood, K., \& Gallagher, J. S. 2011, \apj, 741, 6

\bibitem[Matthews et al.(1999)]{matthews99}Matthews, L. D., Gallagher, J. S., \& van Driel, W. 1999, \aj, 118, 2751

\bibitem[Matthews(2000)]{matthews00}Matthews, L.D., 2000, \aj, 120, 1764

\bibitem[Matthews \& van Driel(2000)]{matthews00b}Matthews, L. D. \& van Driel, W. 2000, \aap, 143, 421

\bibitem[Matthews \& Wood(2001)]{matthews01}Matthews, L. D. \& Wood, K. 2001, \apj, 548, 150

\bibitem[Mo et al.(1994)]{mo94}Mo, H. J., McGaugh, S. S., \& Bothun, G. D. 1994, \mnras, 267, 129

\bibitem[Mosenkov et al.(2010)]{mosenkov10} 
Mosenkov, A.V., Sotnikova, N.Ya., \& Reshetnikov, V.P. 2010, \mnras, 401, 559

\bibitem[Mosenkov et al.(2015)]{mosenkov15} 
Mosenkov, A.V., Sotnikova, N.Ya., Reshetnikov, V.P., et al., 2015, \mnras, 451, 2376

\bibitem[O'Neil et al.(2000)]{oneil00}
O'Neil, K., Bothun, G. D., \& Schombert, J. 2000, \aj, 119, 136

\bibitem[Peebles (1969)]{peebles69}  Peebles, P. J. E. 1969, \apj, 155, 393

\bibitem[Phillips et al.(1990)]{phillips90} 
Phillipps, S., Edmunds, M. G., \& Davies, J. 1990, \mnras, 244, 168

\bibitem[Polyachenko \& Shukhman(1977)]{polyachenko77}
Polyachenko, V. L. \& Shukhman, I. G. 1977, Astronomy Letters, 3, 134

\bibitem[Polyachenko et al.(1997)]{polyachenko97} Polyachenko, V., Polyachenko, E., \& Strel'nikov, A. 1997,
Astron. Letters 23, 525

\bibitem[Rafikov(2001)]{rafikov01} Rafikov, R. 2001, \mnras, 323, 445

\bibitem[Romeo \& Falstad(2013)]{romeo13}
Romeo, A. \& Falstad, N. 2013, \mnras,  433, 1389

\bibitem[Rosenbaum \& Bomans(2004)]{rosenbaum04} 
Rosenbaum, S. D. \& Bomans, D. J.  2004, \aap, 422, 5

\bibitem[Safronov(1960)]{safronov60} Safronov, V. S. 1960, Ann. d'Astrophysique, 23, 979

\bibitem[Saha(2014)]{saha14} 
Saha, K. 2014, arXiv:1403.1711


\bibitem[Schlegel et al.(1998)]{schlegel98} Schlegel, D.J., Finkbeiner, D.P. \& Davis, M., 1998, \apj, 500, 525

\bibitem[Skillman(1987)]{skillman87} 
Skillman, E. D. 1987, Proc. of a conference at the California Institute of Technology, Pasadena, 
California, June 16-19, 1986. Ed. C. J. Lonsdale Persson, p. 263

\bibitem[Sotnikova \& Rodionov(2006)]{sotnikova05} 
Sotnikova, N. \& Rodionov, S. 2005, Astronomy Letters, 31, 17

\bibitem[Sotnikova \& Rodionov(2006)]{sotnikova06} 
Sotnikova, N. \& Rodionov, S. 2006, Astronomy Letters, 32, 649

\bibitem[\protect\citeauthoryear{Spitzer}{1942}]{spitzer42}
Spitzer, L., 1942, \apj, 95, 329

\bibitem[Tempel et al.(2014)]{tempel14} 
Tempel, E., Stoica, R., Mart'nez, V., et al. 2014, \mnras, 438, 3465

\bibitem[Toomre(1964)]{toomre64} 
Toomre, A. 1964, \apj, 139, 1217

\bibitem[Uson \& Matthews(2003)]{uson03} 
Uson, J. \& Matthews, L. 2003, \aj,125, 2455

\bibitem[van der Hulst et al.(1987)]{vdhulst87} 
van der Hulst, J. M., Skillman, E. D., Kennicutt, R. C., \& Bothun, G. D. 1987, \aap, 177, 63

\bibitem[Varela et al.(2012)]{varela12} 
Varela, J., Betancort-Rijo, J., Trujillo, I., \& Ricciardelli, E. 2012, \apj, 744, 82

\bibitem[Vorobyov et al.(2009)]{vorobyov09} Vorobyov, E.I, Shchekinov, Yu., Bizyaev, D., Bomans, D. et al. 2009, \aap, 505, 483

\bibitem[Vorontsov-Velyaminov(1967)]{VV67}Vorontsov-Velyaminov, B. 1967, in Modern Astrophysics, ed. M. Hack (Paris: Gauthier-Villars), p.347

\bibitem[Walker et al.(1996)]{walker96} 
Walker, I. R., Mihos, J. C., \& Hernquist, L. 1996, ApJ, 460, 121


\bibitem[Wright et al.(2010)]{WISE}
Wright, E. L., Eisenhardt, P. R. M., Mainzer, A. K. et al.  2010, \aj, 140, 1868

\bibitem[Zasov et al.(1991)]{zasov91} 
Zasov, A.V., Makarov, D. I., \& Mikhailova, E. A. 1991, Astronomy Letters, 17, 374

\bibitem[Zasov et al.(2002)]{zasov02} 
Zasov, A.V., Bizyaev, D.V., Makarov, D. I., \& Tiurina, N.V., Astronomy Letters, 2002, 28, 527

\bibitem[Zasov \& Khoperskov(2003)]{zasov03} 
Zasov, A. V. \& Khoperskov, A. V., 2003, Astronomy Letters, 29, 437

\end{thebibliography}
\end{document}